\documentclass[conference]{IEEEtran}

\usepackage{amsmath}
\usepackage{amsfonts,amssymb}
\usepackage{bbm}
\usepackage{dsfont}
\usepackage{nicefrac}

\usepackage{multirow}
\usepackage{multicol}
\usepackage{booktabs}
\usepackage{makecell}
\usepackage{tabularx}

\usepackage{graphicx}
\usepackage{subfigure}
\usepackage{float}

\usepackage{xcolor,soul}

\usepackage{algpseudocode}
\usepackage[ruled,vlined,noend]{algorithm2e}

\usepackage{cite}

\usepackage{comment}
\usepackage{acronym}
\usepackage{fancyhdr,lipsum}
\usepackage{microtype}

\SetKwInput{KwInput}{Input}                
\SetKwInput{KwOutput}{Output}              

\fancypagestyle{mahmood}{%
   \fancyhf{} 
   
   \fancyhead[C]{\tiny This manuscript has been accepted for presentation in \textbf{IEEE MECOM 2026}. 
   
   You can use this material personally. Reprinting or republishing this material for the purpose of advertising or promotion, creating new collective works, 
   
   reselling or redistributing to servers or lists, or using any copyrighted component in other works \textbf{must adhere to IEEE policy}. DOI: TBA.}
}%

\begin{document}
\title{Exploiting Overlapping Fields of View for Redundancy-Aware Uplink Transmission in Vehicular 6G}

\author{
    \IEEEauthorblockN{
        Hamidreza Mazandarani\textsuperscript{1}, Masoud Shokrnezhad\textsuperscript{2}, Tarik Taleb\textsuperscript{1}, Onur Günlü\textsuperscript{3, 4} \\
    }
    \IEEEauthorblockA{
       \textsuperscript{1} \textit{Ruhr University Bochum, Bochum, Germany; \{hamidreza.mazandarani, tarik.taleb\}@rub.de} \\
       \textsuperscript{2} \textit{ICTFicial Oy, Espoo, Finland; masoud.shokrnezhad@ictficial.com} \\
       \textsuperscript{3} \textit{TU Dortmund, Dortmund, Germany; onur.guenlue@tu-dortmund.de} \\
       \textsuperscript{4} \textit{Information Theory and Security Laboratory, Linköping University, Linköping, Sweden}
    }
}

\maketitle
\thispagestyle{mahmood}

\begin{abstract}
Emerging uplink-dominant 6G use cases, such as cooperative vehicular streaming, require efficient transmission of high-volume visual data over limited wireless resources. While semantic communications can reduce traffic by prioritizing task-relevant content, most existing approaches treat users independently and therefore overlook spatial redundancy among nearby devices' observations. This paper proposes a semantic-aware multiple access scheme that exploits overlapping fields of view among vehicular users to reduce redundant uplink transmissions. We formulate a joint perception and transmission control problem in which users decide which image patches to transmit, when to transmit them, and over which channel, subject to communication constraints. To address the resulting complexity, we introduce a practical two-phase approach. First, nearby vehicles share selected observation patches over Vehicle-to-Vehicle (V2V) links to calculate inter-user spatial redundancy. Second, users transmit only semantically important, non-redundant patches to the base station, where observations can be reconstructed using the received patches and complementary views from neighboring vehicles. Simulation results in a dense urban vehicular scenario demonstrate that our approach improves the proportion of users who achieve high-fidelity reconstruction, highlighting the potential of semantic-aware multiple access for sustainable and resource-efficient 6G uplink systems.
\end{abstract}

\begin{IEEEkeywords}
6G, Semantic-awareness, Semantic Communications, Resource Allocation, Multiple Access, Medium Access Control (MAC), Wireless Spectrum, generative AI, collaborative wireless networking, vehicular networks.
\end{IEEEkeywords}

\section{Introduction}

The phrase ``Not all bits are equal'' captures the guiding principle behind the semantic revolution in communications, where the significance of source bits is evaluated from the receiver's perspective and leveraged to manage various network functionalities, including multiple access to the frequency spectrum \cite{10580932, 10570768}. Recent advancements in the Artificial Intelligence (AI) domain have facilitated the extraction of non-linear patterns from data, enabling the identification of semantic segments (meaningful data regions) and the prioritization of their transmission according to importance to the downstream task \cite{10521803}. For instance, in a vehicular uplink streaming service with front-view cameras mounted on vehicles, a safety application may focus exclusively on segments that contain pedestrians, pets, and cyclists, whereas a navigation system may prioritize road and traffic sign information. For each of these applications, prioritized segment transmission can be realized within the framework of semantic communications \cite{11370276}.

Nonetheless, most existing literature focuses on either single-user settings or scenarios involving multiple independent users, where the significance of each bit is only relevant within an individual user’s context \cite{11369909, 11175620, 10845862, men2026videotokencom, 11358967}. For instance, Devoto \textit{et al.} \cite{11369909} introduce a transformer-based framework that adaptively identifies and transmits only the visual tokens most relevant to the target task for edge inference, while dynamically responding to channel fluctuations. Liu \textit{et al.} \cite{11175620} develop a Large Language Model (LLM)-driven agentic gating mechanism within a Mixture-of-Experts (MoE) architecture to determine both the semantic content to be transmitted and the suitable encoder for the given context. In video analytics, Fang \textit{et al.} \cite{10845862} present the Prioritized Information Bottleneck (PIB) framework, which leverages Signal-to-Noise Ratio (SNR) and region-of-interest information to selectively communicate task-critical features from multi-camera video streams. Men \textit{et al.} \cite{men2026videotokencom} propose Video TokenCom, a text-intent-guided framework that encodes videos as discrete tokens, allocates higher source-coding precision to tokens corresponding to user-relevant semantic regions. Lastly, Yao \textit{et al.} \cite{11358967} introduce You Only Transmit Once (YOTO), a semantic communications framework that jointly compresses visual features and enables both image reconstruction and caption generation from a single transmitted representation.

\begin{figure}[t!]
\centerline{\includegraphics[width=2.7in]{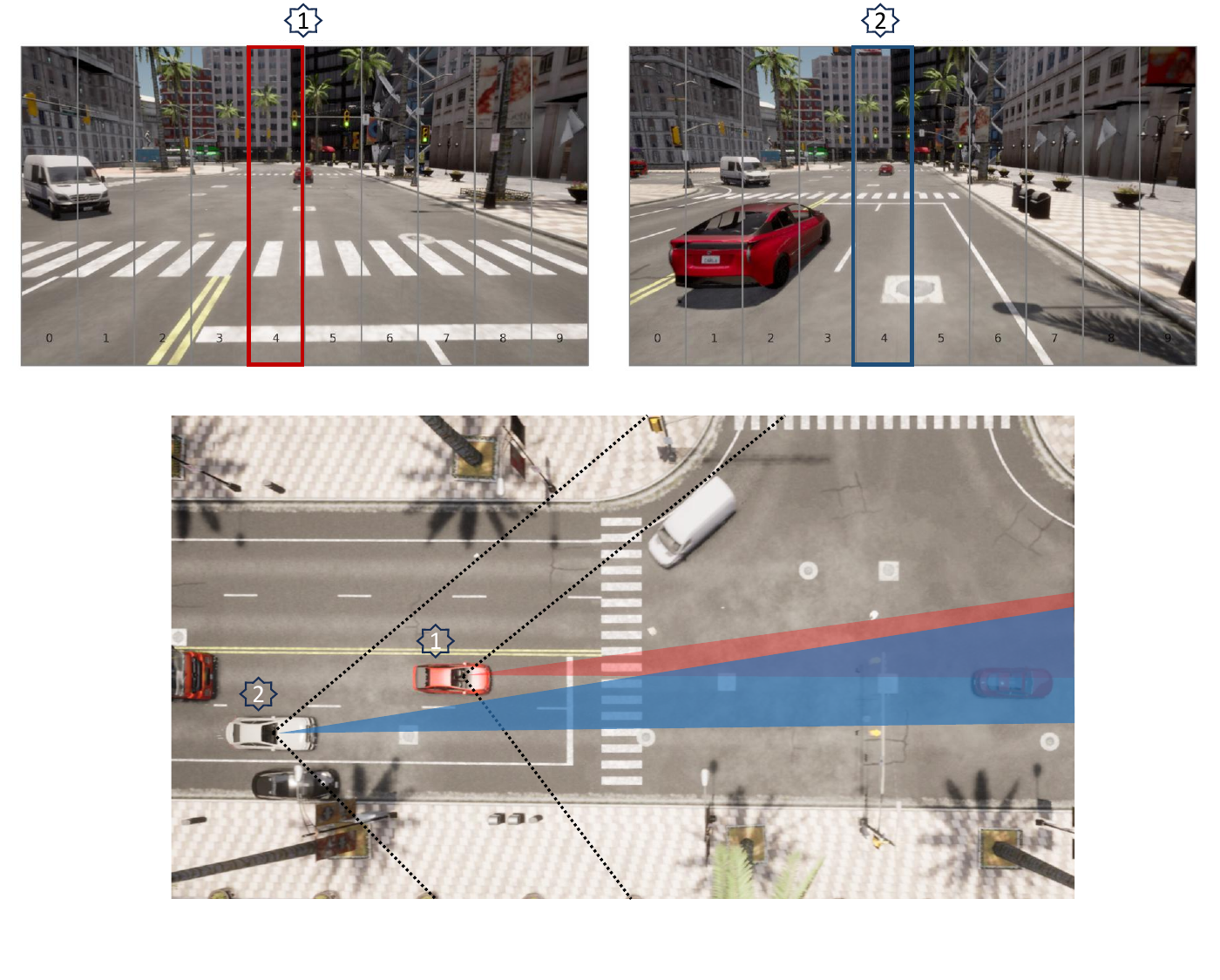}}
\vspace{-15pt}
\caption{Collaborative uplink transmission among nearby vehicles with partially overlapping Fields of View (FoVs), shown by dashed lines. Observation bars are mapped to FoV beams to estimate overlap. For example, the highlighted 4th bars of the red and white vehicles indicate high similarity. Such redundancy can be exploited at the base station to reconstruct one observation using another, reducing redundant uplink transmissions. Viewpoint differences, however, may cause occlusions and object-visibility mismatches, such as traffic lights visible from only one vehicle.}
\label{fig_sample}
\end{figure}

Although the reviewed studies have made significant contributions, they largely overlook the issue of redundancy among users during simultaneous transmissions. In the vehicular streaming example, consider that traffic sign information useful for navigation is already transmitted by another user. In such a case, transmitting the same information again may be redundant and of no additional value to the receiver. Such scenarios have been studied through the lens of correlated sources \cite{11159236} and, more practically, in wireless sensor networks with sensors covering correlated data sources \cite{8335350}. Nevertheless, prior to the emergence of generative AI (GenAI) methods with the capability for reconstructing data from partially-missing information, extending this paradigm to applications requiring high fidelity was not practical. For the previous example, if the objective is merely to identify traffic signs, redundancy detection is relatively straightforward. However, if the goal is high-fidelity stream display at the receiver, the problem becomes significantly more challenging and requires GenAI to reconstruct data from a partially similar one. For example, as illustrated in Fig. \ref{fig_sample}, the viewpoint of the preceding vehicle (red) partially overlaps with the viewpoint of the following vehicle (white). By leveraging the red-colored vehicle's visual observations and aligning them with its own viewpoint, the white-colored vehicle can avoid transmitting redundant visual information already available through the other user's observations.

Within this emerging communication paradigm, algorithms for multiple access to the frequency spectrum should no longer be viewed solely as communication protocols, but rather as perception control systems. In addition to determining \textit{when} to transmit and over \textit{which channel} in multi-channel settings, multiple access algorithms also decide \textit{what} portions of the source data should be transmitted. To achieve this, such algorithms must acquire a degree of cognitive capability to identify semantically non-redundant data segments. However, this objective is challenging, as the data of other users is not available a priori. One approach to address this issue involves enabling observation sharing among users, in which users employ AI-based methods to compare their observations with those of potentially overlapping users. However, this approach introduces a new challenge in terms of network resource consumption, as sharing observations incurs additional overhead. Adding to the problem, uncompressed high-bandwidth data transmission is becoming increasingly important due to the stringent latency and sustainability demands of emerging services, where per-frame compression is not always feasible \cite{11208841}. Thus, this paper focuses on uplink-dominant 6G use cases, particularly vehicular streaming over constrained network resources, by exploiting redundancy in user observations. To address this challenge, we first introduce a novel problem formulation that jointly captures perception and transmission control. We then propose a practical real-time approach in which users share their observations, identify redundant data segments, and selectively transmit the most unique parts to the Base Station (BS) based on the available bandwidth. The BS can then reconstruct the observations with high fidelity using advanced GenAI algorithms, such as conditional diffusion models, Variational Autoencoders (VAEs), and transformer-based sequence models \cite{10422716}.

The rest of this paper is organized as follows. Section \ref{sec_system_model} presents the proposed system model for joint observation sharing and uplink transmissions. Section \ref{sec_approach} provides a practical approach to solving the proposed problem. Section \ref{sec_evaluation} provides performance evaluation results. Finally, Section \ref{sec_conclusion} concludes the paper.

\section{System model and Problem Formulation} \label{sec_system_model}

\subsection{System Model}

\begin{figure}[!t]
\centerline{\includegraphics[width=2.5in]{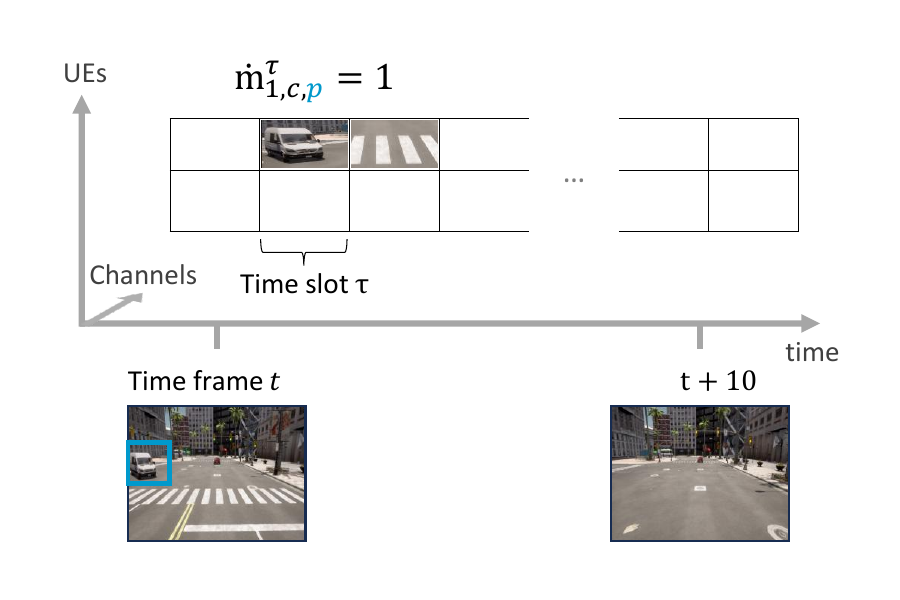}}
\vspace{-15pt}
\caption{Temporal structure of the system model. Each time frame is aligned with the camera frame rate, while each time slot corresponds to a multiple-access decision. In the illustrated example, the first vehicle in Fig.~\ref{fig_sample} is selected to transmit the blue patch.}
\label{fig_time_structure}
\end{figure}

We consider a set $\mathbb{V}$ of vehicular UEs navigating in a small-scale environment (e.g., a city zone) with a limited frequency spectrum for transmitting their camera observations to the BS. While the vehicles' signals interfere with each other, their information remains mutually helpful. Specifically, we consider an environment, where each UE $i$ maintains an approximately constant location ${l}^{t}_{i} \in [0, W]^2$ during \textit{time frame} $t \in \mathbb{T} = \{0, \ldots, \mathcal{T} \}$ and observes a scene denoted with ${o}^{t}_{i}$. Each time frame $t$ consists of $|\mathfrak{T}_{t}|$ communication \textit{time slot}s. At the start of each new time frame, the UE begins transmitting the current observation. At the same time, the receiver reconstructs the previous observation based on the data accumulated up to that time (Fig. \ref{fig_time_structure}).

Each observation ${o}^{t}_{i}$ is partitioned into a set of rectangular patches indexed by $\mathbb{K} = \{1, \ldots, \mathcal{K} \}$, forming the patch sequence $\vec{\bold{o}_{i}^{t}} = \left( {o}^{t}_{i, 1}, \ldots, {o}^{t}_{i, \mathcal{K}} \right)$. A subset of channels, denoted by $\mathbb{C}_{\mathrm{V2V}} \subseteq \mathbb{C}$, is allocated for low-power, high-bandwidth directed transmissions of patches between vehicles. The remaining channels, given by $\mathbb{C}_{\mathrm{V2I}} = \mathbb{C} \setminus \mathbb{C}_{\mathrm{V2V}}$, are used for transmitting patches to the BS. All channels are associated with a collision clique set $\psi^{\tau}_{c} \in \Psi$, which specifies a subset of users that mutually interfere. However, the clique sizes may differ between V2V and Vehicle-to-Infrastructure (V2I) channels due to their distinct transmission ranges and interference characteristics.


\subsection{Problem Formulation}

The problem defined in this section optimizes a policy that, based on the available, limited information, decides which patch to transmit over which channel and in which time slot. The binary decision variable is the transmission policy $\dot{\pi}_{i, c, p}^{\tau}$, which maps the user observation ${o}^{t}_{i}$ and local information $\chi^{\tau}_{i}$ (to be explained later) to transmission decisions represented by the matrix $\dot{\mathbb{M}} = [\dot{m}_{i, c, p}^{\tau}]_{\mathcal{V} \times \mathcal{C} \times \mathcal{K} \times (\mathcal{T} \cdot |\mathfrak{T}|) }$, such that $\dot{m}_{i, c, p}^{\tau}$ indicates whether user $i$ transmits patch $p$ on channel $c$ at the start of time slot $\tau$, as follows: 

\vspace{-8pt}
\begin{equation}
\dot{m}_{i, c, p}^{\tau} = \dot{\pi}_{i, c, p}^{\tau}({o}^{t}_{i}, \chi^{\tau}_{i}) \label{policy}
\end{equation}

The maximum number of patches that can be transmitted in a given time slot depends on factors, such as stream quality and channel bandwidth, denoted with $\mathfrak{B}_{c}$\footnote{In a typical 5G system, each time slot can carry on the order of tens of kilobits. If a frame of several megabits is divided into 100 patches (e.g., a $10 \times 10$ grid), a single time slot can accommodate a few patches. }. The following constraint captures this bandwidth limitation.

\vspace{-10pt}
\begin{equation} \label{bandwidth}
    \sum_{p \in \mathbb{K}}^{}{\dot{m}_{i, c, p}^{\tau}} \le \mathfrak{B}_{c} \quad \forall i \in \mathbb{V}, \forall c \in \mathbb{C}, \forall \tau \in \bigcup_{t \in \mathbb{T} }^{}{\mathfrak{T}_{t}}
\end{equation}
\vspace{-5pt}

Moreover, we next enforce that each user transmits any given patch at most once within a time frame for each channel group. This constraint is justified, as physical constraints to be explained later inherently prevent collisions, and there is no point in duplicate transmissions.

\begin{equation} \label{no_patch_duplicate_transmit}
    {z}^{t}_{i, g, p} = \sum_{c \in \mathbb{C}_{g}, \tau \in \mathfrak{T}_{t}} \dot{m}_{i, c, p}^{\tau} \leq 1 \quad \substack{\forall g \in \{ {\mathrm{V2I}}, {\mathrm{V2V}} \}, \\ \\ i \in \mathbb{V}, p \in \mathbb{K}, \forall t \in \mathbb{T}},
\end{equation}

where the auxiliary variable $z$ indicates whether a patch has been transmitted during the time frame, through V2V links or via a V2I link.

The binary variable $\bar{m}_{i, c}^{\tau}$ indicates whether user $i$ has transmitted at least one patch in time slot $\tau$ on channel $c$, such that we have:

{\footnotesize
\begin{align}
&\big({\sum_{p \in \mathbb{K}}^{}{\dot{m}_{i, c, p}^{\tau}}}\big) / {\mathfrak{B}_{c}} \leq \bar{m}_{i, c}^{\tau} \leq {\sum_{p \in \mathbb{K}}^{}{\dot{m}_{i, c, p}^{\tau}}} \notag \\
&\quad \forall i \in \mathbb{V}, c \in \mathbb{C}, \forall \tau \in \bigcup_{t \in \mathbb{T} }^{}{\mathfrak{T}_{t}} \label{m_bar}
\end{align}
}

Physical constraints imposed below ensure that each user can transmit on at most one channel per time slot due to hardware limitations, and no more than one mutually interfering user can transmit simultaneously on the same channel.

{\footnotesize
\begin{align}
&\sum_{c \in \mathbb{C}} \bar{m}_{i, c}^{\tau} \leq 1 \quad \forall i \in \mathbb{V}, \forall \tau \in \bigcup_{t \in \mathbb{T} }^{}{\mathfrak{T}_{t}} \label{max_channel_constraints} \\
&\sum_{i \in \psi^{\tau}_{c}} \bar{m}_{i, c}^{\tau} \leq 1 \quad \forall \psi^{\tau}_{c} \in \Psi, c \in \mathbb{C}, \forall \tau \in \bigcup_{t \in \mathbb{T} }^{}{\mathfrak{T}_{t}} \label{max_user_constraints}
\end{align}
}

The policy optimized for this problem must detect redundant patches to better manage time and channel resources. We refer to the causal data required for decision-making as \emph{local information}, defined as the set of all observations transmitted by all users before time slot $\tau$ and received by user $i$. This information is denoted by $\chi_i^{\tau}$ and is computed as follows, where $A_{i,j}$ indicates the existence of a V2V link between users $i$ and $j$.

{\footnotesize
\begin{align}
\chi_i^\tau 
= \left\{ {o}^{t}_{j, p} \;\middle|\; \forall j \in \mathbb{V},\; \forall p \in \mathbb{K}:\; \sum_{\substack{c \in \mathbb{C}_{\mathrm{V2V}} \\ \tau' \in \mathfrak{T}_t,\; \tau' < \tau}} A_{i,j} \cdot \dot{m}_{j, c, p}^{\tau'} > 0
\right\} \label{all_observations}
\end{align}
}

At the end of each time frame, the BS concatenates the transmitted observation patches from all UEs to reconstruct the observation for each UE, denoted by $\tilde{o}^{t}_{i}$.

\begin{equation}
\tilde{o}^{t}_{i} = \texttt{Concatenate}\big( \{o^{t}_{j \to i, p} \ | \ z^{t}_{j, \small\mathrm{V2I}, p'} = 1\} \big), \label{Concatenate}
\end{equation}

where $o^{t}_{j \to i,p}$ denotes patch $p'$ from UE $j$, transformed into the coordinate system or viewpoint of UE $i$, with $p'$ matched to patch $p$ in UE $i$. This transformation is obtained using advanced GenAI-based approaches for multi-view synthesis. Notably, when $j=i$, no transformation is required, and thus $o^{t}_{i \to i,p} = o^{t}_{i,p}$.

Finally, we define the objective function as the number of UEs whose reconstruction quality ($\mathcal{Q}^{t}_{i}$) exceeds a predefined threshold. This quality can be determined using structural similarity metrics (e.g., SSIM), perceptual metrics (e.g., Learned Perceptual Image Patch Similarity (LPIPS)), or a combination of both, depending on the target task.

\vspace{-10pt}
\begin{align} \label{problem}
    &\max_{\dot{\Pi}} \ \sum_{\forall t \in \mathbb{T}, \forall i \in \mathbb{V}}^{} \mathds{1} \big(\mathcal{Q}(\tilde{o}^{t}_{i}) \ge Q_{th}\Big) \tag{$P$} \\
    &\mbox{s.t.
    \eqref{policy}, \eqref{bandwidth},  \eqref{no_patch_duplicate_transmit}, \eqref{m_bar}, \eqref{max_channel_constraints}, \eqref{max_user_constraints}, \eqref{all_observations}, \eqref{Concatenate}} \notag
\end{align}

Problem (P) is highly challenging to solve due to the vision-based operations required for view transformation and user-quality evaluation. In addition, an offline optimal solution is generally noncausal, since it would require knowledge of future observations that is not available in real time. Furthermore, the high dimensionality of the joint decision space motivates us to decompose the problem into separate patch-selection and MAC subproblems, as described in the next section.




\section{Proposed Approach} \label{sec_approach}

\begin{figure}[!t]
\centerline{\includegraphics[width=3.5in]{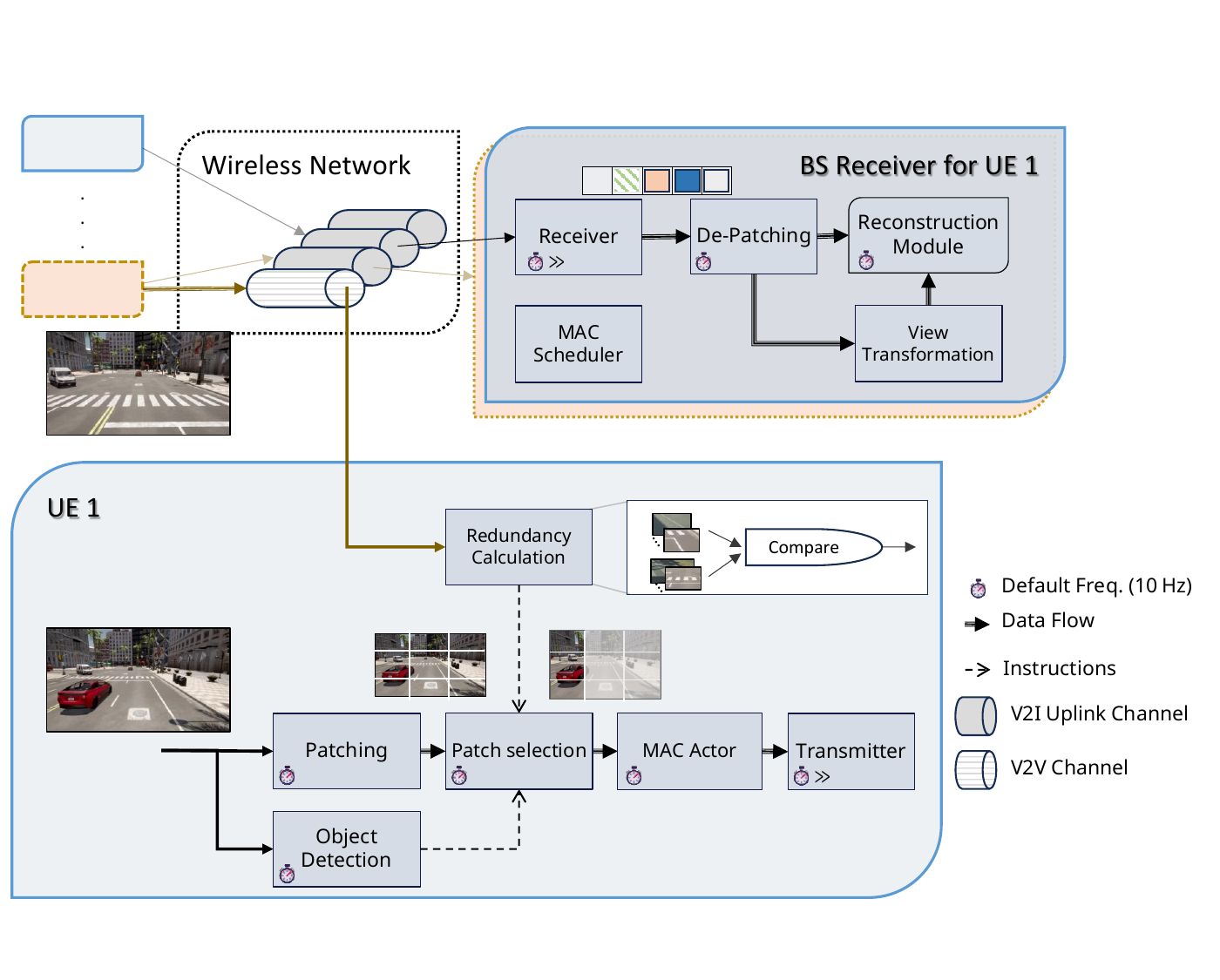}}
\vspace{-20pt}
\caption{High-level overview of the proposed approach pipeline. UEs exploit V2V-shared observations to estimate spatial redundancy, identify informative frame patches, and transmit them to the BS, where the received observations are reconstructed. Patch selection is guided by both the estimated redundancy levels and the presence of detected objects of interest.}
\label{fig_system_model}
\end{figure}

\vspace{-3pt}
To address the aforementioned challenges arising from problem complexity, nonlinearity, and incomplete knowledge of patch similarities, we propose a semantic-aware resource allocation algorithm that performs channel allocation and selects the subset of patches to be transmitted by all users. As summarized in Algorithm \ref{alg}, our approach consists of two phases in each time frame at the transmitter side: first, observation sharing among nearby vehicles over the V2V channel; and second, the actual uplink transmission to the BS\footnote{Our problem formulation does not require separate V2V and V2I transmission phases. However, we adopt this structure for ease of implementation.}. These two phases are described in the following sections. In addition, a high-level overview of the proposed pipeline is illustrated in Fig.~\ref{fig_system_model}.

In the dense vehicular environment considered in this work, each vehicle is likely to have at least one nearby vehicle with a partially overlapping FoV. We refer to the vehicle ahead as the \textit{preceding} vehicle and the vehicle behind as the \textit{following} vehicle. Consequently, multiple preceding-following pairs may coexist in the network, although some vehicles may not belong to any pair. Furthermore, a vehicle may simultaneously serve as the preceding vehicle for one neighbor and the following vehicle for another. For simplicity in the design and evaluation of the proposed transmission schemes, we assign a single role to each vehicle in each frame. These roles may vary over time due to speed variations, lane changes, and overtaking maneuvers. In our implementations, a following vehicle is one located behind the preceding vehicle, within a specified distance threshold (e.g., 50 meters) and a specified orientation threshold (e.g., 90 degrees).

\subsection{Transmitter-side Procedures}

\textbf{Observation Sharing Phase}: At the beginning of each time frame, preceding vehicles start to share their observations so that following vehicles can determine which sets of patches are similar. Since only the V2V channel is utilized in this phase, the constraint in Eq. \eqref{max_channel_constraints} is active, while no collision constraint is assumed to be imposed by Eq. \eqref{max_user_constraints}. The duration of the Observation Sharing (OS) phase is $D_{OS} \leq |\mathfrak{T}|$, which the BS should estimate based on the relative locations and angles of the vehicles or a geometrical technique. Note that the BS does not have access to the source data. Therefore, it must estimate the required time for this phase.


The next challenge is to determine how to prioritize patches for exchange so that the patches most likely to be similar are available to vehicle pairs for comparison. As depicted in Fig.~\ref{fig_sample}, we use a geometric technique (expressed as \textit{FoV-Priority()} in Algorithm \ref{alg} given below) that maps each vertical bar in a vehicle's observation to a corresponding segment of the vehicle's planar FoV. The preliminary similarity scores of the patches belonging to a given vertical bar are then estimated by the overlapping area of the corresponding FoV segments. Although this approach is not fully accurate, making observation sharing relevant, it provides a prior estimate for prioritizing patch sharing.

\textbf{Data Transmission Phase} Once the selected patches are shared among UEs (i.e., local information), they perform pairwise similarity comparisons across all available pairs of patches according to a similarity metric. Rather than restricting the comparison to equally indexed patches, all possible patch combinations are considered, as variations in viewing angles may cause the most similar regions to reside at different indices. Our evaluations show that structural similarity metrics, such as Structural Similarity Index Measure (SSIM), perform better when the shared data are transformed into the subject UE’s viewpoint using GenAI-based methods. Alternatively, patch encodings, such as JEPA embeddings, can be compared using cosine similarity, particularly to identify common semantic elements across two patches. The UEs then perform one-to-one patch matching using the Hungarian algorithm, implemented via the linear sum assignment method, to find the assignment that maximizes the total similarity between the two sets of patches. Finally, each UE selects for transmission only the subset of patches whose matched similarity is below a predefined threshold, as these patches are likely to contain unique information. In Algorithm \ref{alg}, the process that results in the selection of a subset of patches is represented by \textit{PatchSelection()}. However, due to limited frequency resources, not all UEs can transmit all of their selected patches (denoted with $\mathbb{Z}_{\textit{init}}$) during the data transmission phase. Therefore, the BS performs multiple-access resource allocation and schedules the UEs for the data transmissions.

\subsection{Receiver-side Procedures}


\textbf{Resource Scheduling} We adopt a centralized scheduling mechanism in which the BS allocates time slots within each frame across the available channels based on the semantic importance of each UE’s data. This importance is primarily determined by the degree of overlap with target objects and the estimated depth, or alternatively by depth measurements obtained from a depth camera. In our implementation, the target object set includes vehicles, pedestrians, bicycles, traffic lights and signs, lane markings, and crosswalks; however, this set can be adapted according to application-specific requirements. Moreover, the boundaries of salient objects are extracted from the CLIPSeg model's output \cite{Luddecke_2022_CVPR}.

In addition to individual semantic importance, higher priority should be assigned to patches that are expected to contribute to the reconstruction of other users' data, since transmitting such patches can benefit two UEs simultaneously. Accordingly, the importance of ${o}^{t}_{i, p}$ (i.e., patch $p$ of UE $i$ at time frame $t$), is denoted by ${\omega}_{i, p}^{t}$ and is computed using:

\begin{align} \label{score}
{\omega}_{i,p}^{t}
=
\bar{c}_{i,p}^{t}
\left(1+\beta \cdot ( \frac{d_{\max}-\bar{d}_{i,p}^{t}}{d_{\max}-d_{\min}} )\right)
\left(1+\gamma \cdot r_{i,p}^{t}\right),
\end{align}
where $\bar{c}_{i,p}^{t}$ represents the average target-object coverage of the patch, $\bar{d}_{i,p}^{t}$ is the average depth value of patch, and $r_{i,p}^{t}$ indicates whether this patch is intended to be used for reconstructing another similar patch. The parameters $\beta$ and $\gamma$ control the relative contributions of depth and redundancy, respectively.

\textbf{Data Reconstruction} At the BS, patches received from all UEs are aggregated, and similar patches are identified using the same procedure employed at the UEs. Alternatively, this metadata can be forwarded by UEs to the BS. Missing patches may be recovered using GenAI-based inpainting, such as transformer-based methods \cite{elharrouss_transformerbased}, but these approaches generate plausible content rather than guaranteed ground truth and may introduce hallucinations. A more reliable alternative is to exploit observations from nearby UEs. For example, RGB-D view transformation can match pixels across views using LoFTR \cite{Sun_2021_CVPR}, associate them with depth, estimate the relative pose, and warp source-view pixels into the target viewpoint. However, geometry-based reconstruction requires accurate calibration and is most effective for static scenes; dynamic objects and occlusions can cause warping artifacts. Alternatively, reasoning-enabled Large Multi-Modal-based reconstruction can use nearby observations, maps, and semantic cues to guide inpainting and plausibility assessment, improving consistency at the cost of higher computation and delay.

\begin{algorithm}[t!]\label{alg}
\small
\caption{\small{Semantic-Aware Multiple Access via Spatial Redundancy Exploitation}}
\KwInput{Observation Sharing phase duration $D_{OS}$, similarity threshold $\eta$, bandwidths $\{\mathfrak{B}_{c}\}_{c \in \mathbb{C}}$}
\KwOutput{Reconstructed observations $\{\tilde{o}_{i}^{t}\}$}

\ForEach{$t$ in $\mathbb{T}$}
{
    $\dot{\mathbb{M}}^{t}, \mathcal{Z}_{\textit{init}}^{t}, \boldsymbol{\omega}^{t} \gets 0, 0, 0$ \\

    \ForEach{$i$ in $\mathbb{V}$}
    {
        \textcolor{gray}{UE $i$:} Partition ${o}_{i}^{t}$ into $\{{o}_{i,p}^{t} \ | \ p \in \mathbb{K}\}$ \\
        \textcolor{gray}{UE $i$:} $\mathbb{K}_{i,\textit{share}}^{t} \gets$ \textit{FoV-Priority}\Big($i, \{l_{i}^{t}\}_{i \in \mathbb{V}}$\Big)
    }

    \ForEach{$\tau$ in $\mathfrak{T}_{t}$}
    {
        \If{$\tau < D_{OS}$}
        {
            \ForEach{$i$ in $\mathbb{V}_{\textit{Preceding}}$}
            {
                \textcolor{gray}{UE $i$:} Transmit next $\mathfrak{B}_{c}$ patches from $\mathbb{K}_{i,\textit{share}}^{t}$ over $c \in \mathbb{C}_{\mathrm{V2V}}$ \\
            }
        }

        \If{$\tau = D_{OS}$}
        {
            \ForEach{$i$ in $\mathbb{V}$}
            {
                \textcolor{gray}{UE $i$:} Build local information $\chi_{i}^{D_{OS}}$ acc. \eqref{all_observations} \\
                \textcolor{gray}{UE $i$:} $\mathbb{Z}_{\textit{init},i}^{t} \gets \textit{PatchSelection}\Big(\vec{\bold{o}_{i}^{t}}, \chi_{i}^{D_{OS}}, \eta$\Big) \\
            }

            \textcolor{gray}{BS:} Calculate $\boldsymbol{\omega}^{t}$ acc. \eqref{score} \\
            \textcolor{gray}{BS:} $\dot{\mathbb{M}}^{t} \gets \textit{Schedule}\Big(D_{OS}, \mathbb{Z}_{\textit{init}}^{t}, \boldsymbol{\omega}^{t}, \{\mathfrak{B}_{c}\}_{c \in \mathbb{C}_{\mathrm{V2I}}}\Big)$ \\
            
        }

        \If{$\tau > D_{OS}$}
        {
            \ForEach{$i, c, p$ in $\mathbb{V}, \mathbb{C}_{\mathrm{V2I}}, \mathbb{K}$}
            {
                \If{$\dot{m}_{i,c,p}^{\tau} = 1$}
                {
                    \textcolor{gray}{UE $i$:} Transmit $o_{i, p}^{t}$ to the BS \\
                }
            }
        }

        \If{$\tau = |\mathfrak{T}_{t}|$}
        {
            \ForEach{$i$ in $\mathbb{V}$}
            {
                \textcolor{gray}{BS (Optional):} Transform available patches into UE $i$ viewpoint \\
                \textcolor{gray}{BS: } Reconstruct $\tilde{o}_{i}^{t}$ acc. \eqref{Concatenate} \\
            }
        }
        
    }

}
\end{algorithm}

\section{Evaluation} \label{sec_evaluation}

To evaluate the performance of our proposed approach, we deploy a dense network of $40$ vehicles navigating an urban zone in the CARLA simulator for $10$ seconds ($10K$ time slots). Each vehicle records RGB camera observations at $10$ frames per second with a resolution of $1280 \times 720$ pixels. Each frame interval is divided into $100$ time slots and consists of two communication phases. In the first phase, a fixed number of slots, e.g., $5$, is allocated for observation sharing among vehicles with the V2V channel bandwidth of $\mathfrak{B}_{V2V} = 10$. In the second phase, the remaining slots are used to transmit image patches to the BS. Each observation frame is partitioned into $100$ patches arranged on a $10 \times 10$ grid. The implementation accounts for practical physical-layer constraints, including single-channel usage by each UE in any time slot (Eq. \eqref{max_channel_constraints}) and collision-free channel allocation (Eq. \eqref{max_user_constraints}). 


We evaluate our algorithms using two reference cases. The first uses the unmodified observation of the preceding user as a lower-bound reference. The second assumes perfect view transformation of the preceding user’s observation into the following user’s viewpoint, which is equivalent to directly accessing the following user’s observation and thus serves as an upper bound. Hence, the performance of any practical reconstruction algorithm is expected to lie between these two cases.

\begin{figure}[!thb]
\centerline{\includegraphics[width=3.7in]{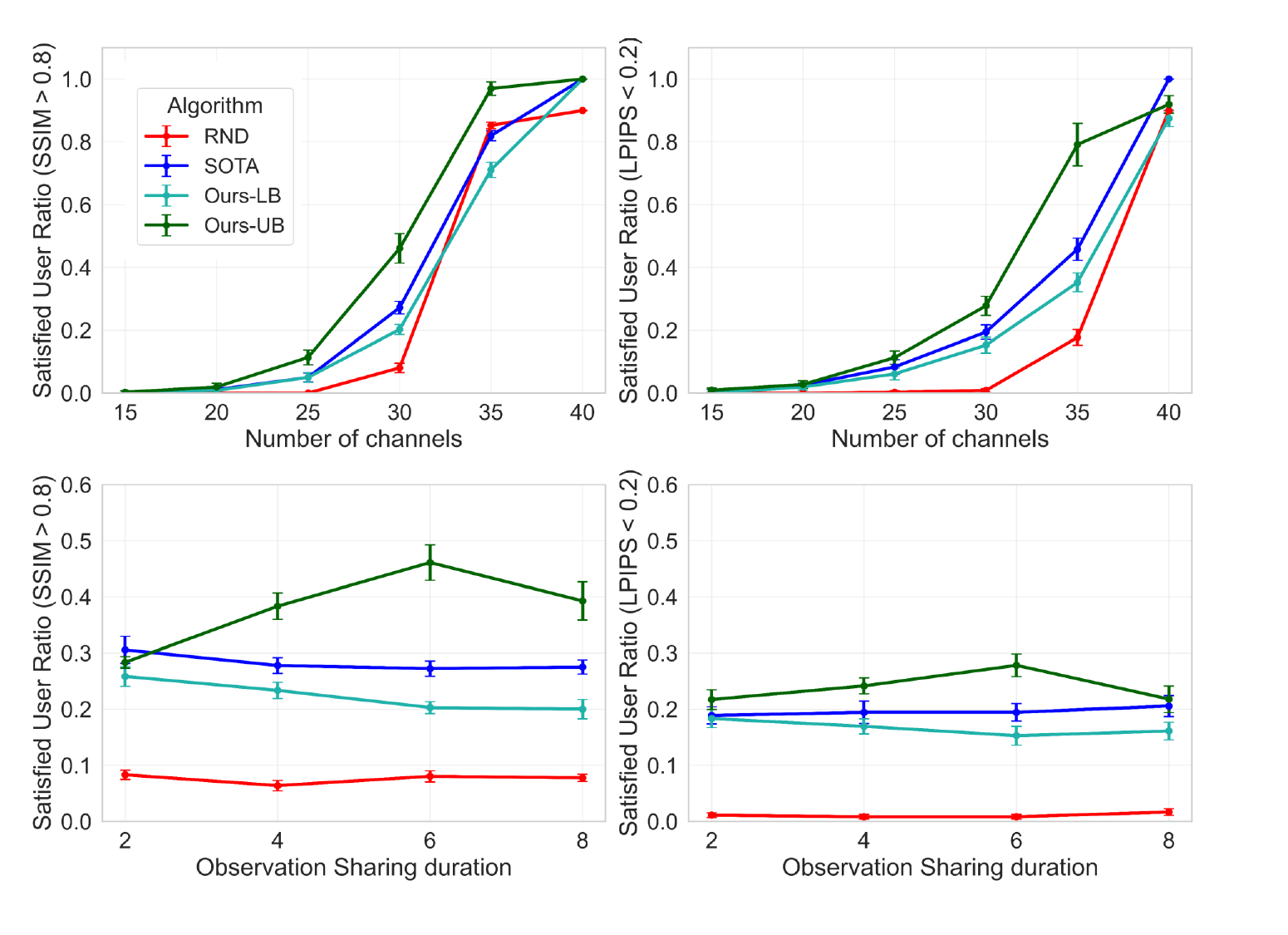}}
\caption{Performance comparison of the centralized semantic-aware state-of-the-art (SOTA) algorithm, Lower Bound (LB) of our approach without view transformation, and Upper Bound (UB) of our approach with perfect view transformation. The top row shows the ratio of satisfied users as a function of the number of available channels, while the bottom row shows performance under different observation sharing periods. SSIM- and LPIPS-based user satisfaction ratios are reported with error bars indicating variability across different time frames and UEs.}
\label{fig_results}
\end{figure}

The results show that the proposed approach with perfect view transformation, denoted by Ours Upper-Bound (UB), consistently outperforms importance-aware semantic transmission baselines such as \cite{10521803}, denoted by SOTA. As expected, increasing the number of available channels improves the performance of all schemes, as shown in the upper part of Fig.~\ref{fig_results}. However, the gain achieved by exploiting spatial redundancy is most pronounced in the moderate-resource regime. In this regime, the available bandwidth is insufficient to transmit all users' observations, yet sufficient for selective transmission to provide meaningful reconstruction quality. Consequently, avoiding redundant transmissions enables more efficient use of uplink resources and improves the ratio of satisfied users.

The lower part of Fig.~\ref{fig_results} evaluates the impact of the observation-sharing duration, i.e., the initial portion of each frame allocated to V2V exchange among nearby vehicles. The results reveal a unimodal trend, highlighting the trade-off between redundancy discovery and uplink transmission time. When the sharing duration is too short, vehicles obtain limited information from neighboring observations, reducing the ability to identify redundant patches. In this case, the proposed method behaves closer to the SOTA baseline. Conversely, when the sharing duration is too long, fewer time slots remain for V2I transmission, which reduces the amount of data delivered to the BS and degrades reconstruction quality. These results indicate that optimizing the observation-sharing duration is critical for balancing cooperation overhead and uplink efficiency.

\vspace{-5pt}
\section{Conclusion} \label{sec_conclusion}

This paper introduced a semantic-aware multiple access framework for uplink-dominant 6G scenarios in which nearby vehicles have partially overlapping visual observations. Unlike conventional MAC schemes, the proposed approach treats multiple access as a joint communication and perception control problem, where users also determine which parts of their observations are worth transmitting. To this end, we formulated an optimization problem that captures patch-level transmission decisions.

To enable real-time operation, we proposed a two-phase procedure: V2V observation sharing for local redundancy estimation, followed by semantic-aware uplink transmission of non-redundant patches. Our evaluations show that exploiting spatial redundancy improves the ratio of users meeting perceptual quality targets, particularly under limited uplink resources.

Future work will integrate practical reconstruction modules, including geometry-aware view synthesis and multimodal generative models. Another direction is to study robustness against dynamic objects, imperfect V2V links, calibration errors, and delayed or partial observation sharing, for which multi-hop deep joint source-channel coding methods, as in \cite{11462344}, can be used.

\section*{Acknowledgment}
The research work is supported in part by the German Federal Ministry of Research, Technology, and Space (BMFTR), through the Project 6GEM+ under Grants 16KIS2411, 16KIS2412, and 16KISS005; and in part by the European Union's Horizon Europe research and innovation programme under the 6G-Path project (Grant No. 101139172).

\bibliographystyle{IEEEtran}
\bibliography{IEEEabrv,main}

\end{document}